\documentstyle[preprint,aps,eqsecnum]{revtex}
\tightenlines

\def\beq#1{\begin{equation} \label{#1}}
\def\eeq{\end{equation}}
\newcommand{\bea}{\begin{eqnarray}}
\newcommand{\eea}{\end{eqnarray}}
\def\bra#1{\left\langle #1\right\vert}
\def\ket#1{\left\vert #1\right\rangle}
\def\epsp{\epsilon^{\prime}}
\def\NPB{{ Nucl. Phys.} B}
\def\PLB{{ Phys. Lett.} B}
\def\PRL{ Phys. Rev. Lett.}
\def\PRD{{ Phys. Rev.} D}

\begin{document}
{
\tighten

\title {Quantum Theory of Neutrino Oscillations for Pedestrians - Simple 
Answers to Confusing Questions}
\author{Harry J. Lipkin}
\address{ \vbox{\vskip 0.truecm}
  Department of Particle Physics
  Weizmann Institute of Science, Rehovot 76100, Israel \\
\vbox{\vskip 0.truecm}
School of Physics and Astronomy,
Raymond and Beverly Sackler Faculty of Exact Sciences,
Tel Aviv University, Tel Aviv, Israel  \\
\vbox{\vskip 0.truecm}
High Energy Physics Division, Argonne National Laboratory,
Argonne, IL 60439-4815, USA\\
~\\harry.lipkin@weizmann.ac.il
\\~\\
}

\maketitle

\begin{abstract}  

A simple rigorous calculation  confirms the standard formula and clarifies some
confusing difficulties arising in the standard textbook recipe converting the
unobserved frequency of time oscillations between neutrino states with
different energies to the observed oscillation wave length in space  Including
the quantum fluctuations in the position of the detector and in the transit
time between source and detector enables the treatment of: 1. The difference in
velocity and transit time between neutrinos with different energies. 2. The
destruction of all phases between states with different masses by an ideal
detector which measures the energy and momentum of the neutrino. 3. The
destruction of all phases between states with different energies by a realistic
detector in thermal equilirium with its macroscopic environment. 4. The
difficulty for relativistic treatments and relativistic field theory to treat
the crucial quantum mechanics of a macroscopic detector at rest in the
laboratory.

\end{abstract}

} 


\def\beq#1{\begin{equation} \label{#1}}
\def\eeq{\end{equation}}
\def\bra#1{\left\langle #1\right\vert}
\def\ket#1{\left\vert #1\right\rangle}
\def\epsp{\epsilon^{\prime}}
\def\NPB{{ Nucl. Phys.} B}
\def\PLB{{ Phys. Lett.} B}
\def\PRL{ Phys. Rev. Lett.}
\def\PRD{{ Phys. Rev.} D}

\section{Introduction} 

\subsection{Problems causing confusion in understanding neutrino oscillations}

Although an extensive review of neutrino oscillations does not mention time
at all\cite{review} and bypasses all confusion, 
the continuing argument about the roles of energy, momentum and time in neutrino
oscillations\cite{neusb} still causes confusion\cite{cutapp}. The ``equal
energy assumption" of one paper\cite{Okun} was criticized by another
\cite{Giunti} while shown to arise naturally from the interaction of the
neutrino with its  environment\cite{Leo}. This interaction is still often
ignored\cite{Giunti2,Giunti3} and the controversy
continues\cite{OkunA,OkunB,OkunC}.

Why are different mass states coherent? What is the correct formula for the
oscillation phase? How can textbook formulas for oscillations in time  describe
experiments which never measure time? How can we treat the different velocities
and different transit times of different mass eigenstates and avoid incorrect
factors of two? How can textbook forumulas which describe coherence between
energy states be justified when Stodolsky's theorem states there is no
coherence between different energies? Is covariant relativistic quantum field
theory necessary to describe neutrino oscillations? How important is the
detector, which is at rest in the laboratory and cannot be Lorentz tranformed
to other frames?

These questions are answered by a simple rigorous calculation which includes
the quantum fluctuations in the position of the detector and in the transit
time between source and detector.  The commonly used standard formula for
neutrino oscillation phases is confirmed.  An ``ideal" detector
which measures precisely the energy and momentum of the neutrino destroys all
phases in the initial wave packet and cannot observe oscillations. A realistic
detector preserves the phase differences between neutrinos having the same
energy and different momenta and confirms the standard formula. Whether phase
differences between neutrinos with different energies are observable or
destroyed by the detector is irrelevant.

The confusion begins with the textbooks
that claim that a neutrino in a mixture of mass eigenstates 
with different energies will oscillate in time.
Nobody measures time. What is measured is distance.  Confusion arises in
converting the period of oscillation in time to a wave length in distance,
combined with the observation that neutrinos with different energies
travel with different velocities and arrive at the detector at different
times. How to treat this is clear to anyone who understands first year
quantum mechanics and wave-particle duality. Confusion arises because too many 
papers on neutrino
oscillations are written by people who don't and double counting leads to
the factor of two.
     
     A second source of confusion is that neutrino oscillations are observable
only if the neutrino detector is able to detect the relative phase of
components in an incident wave packet. Leo Stodolsky\cite{Leo} has pointed out
that any stationary detector is described by a density matrix that is diagonal
in energy and that phases between incident neutrino amplitudes with different
energies are not observable.  Only interference between components with the
SAME ENERGY and different momenta are observable. But since this point is
continuously ignored in the  literature and in conferences it is spelled out
clearly and unambiguously in the present paper. It may be only first year
quantum mechanics, but  unfortunately it seems to be needed.

These problems are illustrated in the section on neutrino mass, mixing and
flavor change in the review of particle physics\cite{PDG} by the Particle Data
Group. 
The section contains this caveat ``The quantum mechanics leading to the same
result is somewhat subtle. To do justice to the physics requires a more refined
treament than the one we have given. Sophisticated treatments continue to yield
new insights". Unfortunately most readers ignore the caveat and a literal
interpretation of the text contributes to the confusion.

The time-dependent neutrino wave function described by eq. (13.4) has a
``Lorentz-invariant phase factor". The text does not note that this phase
factor is not gauge invariant and is not observable; only phase differences are
observable.  Although first year quantum mechanics tells us that the overall
phase of a wave function is not observable, many readers fall into this trap
and take the phase seriously. 

The text considers a  neutrino produced with momentum p; all components
have same momentum. The time-dependent phase is replaced by a
distance-dependent phase by noting that the neutrino is extremely
relativistic and sets $t\approx L$. However, this is not strictly true for
finite mass neutrinos. Neutrinos with the same momentum and different
masses have different velocities and travel the distance $L$ in sufficiently
different times to cause confusion.

This has recently been pointed out to the Particle Data Group by a reader who
noted that the exact relation between $t$ and $L$ reads $L = t\cdot v$, with
$v= p/E$.  Substituting  $L/v = L\cdot E/p$ into the ``Lorentz-invariant phase
factor" he gets $E\cdot t - p\cdot L = E^2\cdot L/p - p^2\cdot L/p = (E^2 -
p^2)\cdot L/p = m^2\cdot L/p$ which differs from the standard result by a
factor of 2.

The text then examines the  relative phases of the mass eigenstates at the
distance $L$ and converts the state to a linear combination of flavor
eigenstates. This assumes that the states with different energies are coherent
and that their relative phase is observable. But Stodolsky has shown\cite{Leo}
that neutrinos with different energies lose all coherence in the interaction
with the detector and their relative phase is not observable. 

Our purpose here is to present a coherent treatment that can be
easily understood by physicists who are unaware of the  ``somewhat subtle
quantum mechanics" needed to understand neutrino oscillations.

If the interaction with the environment is ignored, the problem reduces to the 
propagation of a single noninteracting Dirac fermion, easily solved without 
relativistic field theory. In a real experiment the neutrino passes the
detector as a wave with a finite  length in time and can interact with the
detector any time during this finite interval. These quantum fluctuations in
transit time affect the relative phase observed at the detector between two
neutrino  waves. Confusion and erroneous factors of two arise in incorrectly
treating this  transit time as a well-defined classical variable.
We present here a rigorous treatment showing that the standard relation between
phases of amplitudes having different masses always holds irrespective of
whether states with equal energy, equal momentum or any other combination are
used. Effects of transit time variations are shown to be negligible.   

The observability of these phases depends on the detector. An ideal detector;
e.g. a nucleon at rest, measures the neutrino mass and destroys all coherence
and relative phases. This  ``missing mass" experiment has no oscillations. 
More realistic  detectors preserve the relative phases of amplitudes  having
the same energy and different momenta; the detecting nucleon is confined to a
region of space small compared to the wave length of the observed neutrino
oscillations. The relative amplitudes  for final states containing an $e$,
$\mu$ or $\tau$ are thus determined for each energy. Since these relative
magnitudes do not change appreciably with energy in  realistic experiments, the
flavor output of the detector is determined without needing Stodolsky's
stationary environment assumption\cite{Leo}. Even if a judicious time
measurement  preserves some coherence between different energies, the flavor
output is unchanged.

\subsection{The basic problem}

The source of the confusion is easily seen by examining the state of a
neutrino wave packet incident on a detector: a mixture of states having
different masses, energies and momenta. The phase change in the wave function
of a component of the wave packet in traveling a distance $x$ from the source
to the detector in a time $t$ is  

\beq{phase}  
\phi(x,t) = px -Et  
\end{equation}  

This phase by itself is unobservable, like any overall phase of a wave
function. It is not gauge invariant since a gauge transformation can multiply
the wave function by an arbitrary function of x.

What is gauge invariant and observable is
the phase difference between any  two components of the wave function having
masses $m_1$ and $m_2$    

\beq{shift}  
\delta \phi(12) = \phi_1(x,t) - \phi_2(x,t) = (p_1 -p_2)x -(E_1 -E_2)t =
{{p_1^2-p_2^2}\over{p_1+ p_2}} \cdot x - {{E_1^2-E_2^2}\over{E_1+ E_2}} \cdot t
\end{equation}  
where $p_1$, $p_2$, $E_1$ and $E_2$ are respectively the momenta and energies 
of these two components in the wave packet.   
  
To relate this result (\ref{shift}) to the phase difference actually observed
in  a detector, we set $x$ equal to the distance between the source and the
detector. The problem arises in choosing a value for t. We parametrize the
problem by setting $t$ equal to the time required to traverse the distance x  
using an average group velocity $\bar v$ for the two mass eigenstates,  with a
correction $\delta t$ to describe any discrepancies, including quantum
fluctuations. 

\beq{velocity2}  
\bar v \equiv {{p_1+ p_2}\over{E_1+ E_2}}; ~ ~ ~
t= {x \over \bar v} +  \delta t  = {{E_1+ E_2}\over{p_1+ p_2}}\cdot x +\delta t 
\end{equation}  
 Thus 
\beq{shiftb}  
\delta \phi(12) =
{{p_1^2-p_2^2}\over{p_1+ p_2}} \cdot x - {{E_1^2-E_2^2}\over{E_1+ E_2}} \cdot t
= {{m_2^2-m_1^2}\over{p_1+ p_2}} \cdot x  - {{E_1^2-E_2^2}\over{E_1+ E_2}} 
 \cdot \delta t
\end{equation}  

This gives the standard formula for the phase shift independent of the momentum
and energy values of the two interfering waves with a correction term which
vanishes when the two waves have the same energy.  The physics here is evident.
The relative phase at the same point in space between two amplitudes with
different energies  is proportional to  the product of the energy difference
and time and sensitive to the  exact time when the neutrino is detected. The
$\delta t$ correction is appreciable for the understanding of time behavior.
But a quantitative analysis given below shows the  correction  to the formula
(\ref{shiftb}) to be negligible. This explains why treatments using waves with
the same momentum and different energies give the same answer as those using
the same energy and different momenta. 

The time behavior has caused much confusion in the community and in the
literature. We therefore we investigate in detail its physical meaning and 
quantum fluctuations. 

\section{The elementary quantum mechanics of space and time}

Neutrinos arrive at a detector at distance $x$ from
the source as a coherent linear combination of states having different masses,
energies and momenta. Since the source and absorber are quantum
systems, quantum fluctuations arise in both the distance $x$ and in the time
interval $t$ between the emission from the source and its observation at the
detector. The distance fluctuations can be shown to be small in any experiment
where the positions of the source and absorber are known to a precision much
smaller than the wave length of the oscillation to be measured. The 
nature of the quantum fluctuations in the time interval $t$ are not generally
well understood and are the source of much confusion.  

The neutrino wave function at the detector defines a probability
amplitude for its detection at any time during the arrival of the
wave packet.  The value of the transit time $t$ is thus not
precisely defined and subject to the usual quantum fluctuations of
any variable described by a quantum-mechanical probability amplitude. 

The waves for different mass eigenstates travel with different group
velocities, and their centers may arrive at the detector at different times.
But the detector does not measure the arrival time of a wave packet center. 
Each event has a single value of the time $t$. The flavor output of the
detector is determined by the relative phases of the different components of
the wave packet at that time. The quantum fluctuations in time can be observed
if there is a measurement of the time spectrum of events. 

We now define the quantities necessary to 
evaluate the phase shifts in eq.(\ref{shiftb}). 

The time variable $t$ to use in eq. (\ref{shift}) is defined as the 
interval between the time when the neutrino is detected and a time at the
source when the neutrino is in a flavor eigenstate. This is then subject to the
uncertainty denoted by $\delta t$. The neutrino flavor at the source can be
measured by placing a detector a short distance from the source. This was in
fact the first experiment that showed  that there were different flavors of
neutrinos.

We now check the effect of $\delta t$ for the case of equal momenta
$p_1 = p_2 \equiv p$ in (eq.\ref{shiftb}).
\beq{shiftb3}  
\delta \phi(12) = {{m_2^2-m_1^2}\over{p_1+ p_2}} \cdot \left( x  + 
{{E_1^2-E_2^2}\over{m_1^2-m_2^2}}\cdot \bar v \cdot \delta t\right) =
{{m_2^2-m_1^2}\over{2p}} \cdot x 
\cdot \left(1 + {{\bar v \cdot \delta t}\over x } \right)
\end{equation}  

Since the spatial size of the wave packet, $ \bar v \cdot \delta t$ is  much
smaller than the distance x between source and detector the neglect of $\delta
t$ in eqs. (\ref{shiftb}) and (\ref{shiftb3}) is clearly justified. But $\delta
t$ is also seen to be much larger than the time separation $\delta
t_{wc}$
between the centers of the two mass eigenstate wave packets, which have moved
through the distance $x$ with group velocities $v_1 = p/E_1$ and  $v_2 = p/E_2$
respectively. 

\beq{shiftb4}  
\bar v \cdot \delta t_{wc} = 
\bar v \cdot \left({x \over v _1 }- {x \over v _2 }\right)
\approx  {{m_2^2-m_1^2}\over{2p^2}} \cdot x  << \bar v \cdot \delta t << x 
\end{equation}   
The transit time fluctuations $\delta t$ are seen to be much larger than 
$\delta t_{wc}$ but sufficiently small so that the correction to eq.
(\ref{shiftb}) is negligible.  
The time difference $\delta t_{wc}$  is automatically included in the above
calculation and needs no further correction. Unnecessary additional corrections
motivated by the erroneous classical picure of a particle traveling on a
classical path with a well defined velocity  has led to the spurious factors of
2.

\section {What is observable}

	Whether the phase (\ref{shiftb}) is experimentally observable depends
upon the detector.  

\subsection {An ``ideal" detector}
	Consider an ``ideal" detector which is a nucleon at rest. It absorbs
the incident neutrino and emits a lepton. Since energy and momentum are
conserved, the energy and momentum transfers  from the neutrino to the detector
denoted by $\delta E$ and $\delta p$ are equal respectively to the energy and
momentum of the incident neutrino, denoted by $E_\nu$ and $p_\nu$  

\beq{deltaE}  
\delta E = T_N + E_{lepton}=E_\nu; ~ ~ ~   
 \delta p =p_N + p_{lepton} = p_\nu   
\end{equation}   
where $T_N$ and $p_N$ are the kinetic energy and momentum of the final nucleon,
and $E_{lepton}$ and $p_{lepton}$ are the total energy and momentum  of the
emitted lepton. 

This ``missing mass experiment" determines the neutrino mass  
\beq{mnu}  
m_\nu=\sqrt{E_\nu^2 - p_\nu^2}  
\end{equation}  
Since the detector determines the neutrino mass, waves arriving at the detector
with different masses are not coherent. All phase information is
destroyed in this ``ideal" detector.  

\subsection {A realistic localized detector}
     
     Realistic detectors are not ideal. The localization of the detector
nucleon in a small region of space makes its momentum uncertain. There is no
comparable localization in time.
This asymmetry between energy and momentum in the initial detector state
destroys the apparent symmetry between energy and momentum noted in all
covariant descriptions of neutrino oscillations. A fully covariant description
of any experiment which can be used also to consider detectors moving with
relativistic velocities is not feasible at present. A covariant description
which neglects the quantum mechanics of the neutrino-detector interaction is
neglecting essential physics of all realistic oscillation experiments.

The following section gives a rigorous treatment of the handwaving argument
using the uncertainty principle to relate the   momentum spread of the wave
function to the size of the detector.   The phase between components of the
neutrino wave with the same energy and different masses and momenta is shown to
be observable if quantum fluctuations in the position of the
detector nucleon are very small in comparison with the wave length of the
neutrino oscillations in space produced by the mass difference between neutrino
mass eigenstates.

The simple hand-waving argument for this physics states that the uncertainty
principle and the localization in space of the detector nucleon that absorbs
the neutrino prevents the detector from knowing the difference between
components of the incident neutrino wave packet with slightly different momenta
and the same energy.

\section {A detailed rigorous calculation of the detection process}

The rigorous quantum-mechanical argument notes that the product  $\delta p
\cdot \delta x$ of the quantum fluctuations in the position of the detector
nucleon $\delta x$ and the range of momenta $\delta p$ in relevant neutrino
states having the same energy is a small quantity. Taking the leading terms in
the expansion of the transition matrix elements in powers of $\delta p \cdot
\delta x$ gives the result that the flavor spectrum of the charged leptons
emitted from the detector at a given energy is determined by the relative phase
of the components of the incident neutrino wave packet having the same energy
and different momenta. Whether coherence between amplitudes with different
energies is destroyed, as required by Stodolsky's theorem, or can be preserved by
judicious time measurements is irrelevant to this result.

The detection of the neutrino is a weak interaction described in first order
perturbation theory by transition matrix elements between the initial state of
the neutrino-detector system before the interaction and all possible final
states. 

Consider the transition matrix element between an initial state $\ket{i(E)}$
with energy $E$ of   the entire neutrino - detector system and a final state
$\ket{f(E)}$ of the system of a charged muon and the detector with the same
energy E, where a neutrino $\nu_k$ with energy, mass and momentum $E_\nu$,
$m_k$ and  $\vec P_o + \vec {\delta P_k}$ is detected via the transition 

\beq{weak1}
\nu_k + p \rightarrow \mu^+ + n  
 \eeq
occurring on a proton in the detector.  We express the neutrino momentum
as the sum of the mean momentum $\vec P_o$ of all the neutrinos  
with energy
$E_\nu$ and the difference $\vec {\delta P_k}$ between the momentum of each 
mass eigenstate and the mean momentum,

The transition matrix element depends upon the individual mass eigenstates $k$
only in the momentum difference $\vec {\delta P_k}$ and a factor $c_{k\mu}$ for
each mass eigenstate which is a function of neutrino mixing angles. $c_{k\mu}$
describes the transition amplitude for this mass eigenstate to produce a muon
when it reaches the detector, multiplied by a phase given by the generalization
to the case of three neutrinos of eq. (\ref{shiftb}) with $E_1=E_2$  The
transition matrix element can thus be written in a  factorized form with one
factor $T_o$ independent of the mass $m_k$ of the neutrino eigenstate and a
factor depending on $m_k$. 

\beq{weak2} \bra{f(E)}T \ket{i(E)} =  \sum_k
\bra{f(E)}T_o \cdot c_{k\mu} e^{i\vec {\delta P_k}  \cdot \vec X} \ket{i(E)}  
\eeq   
where $\vec X$ denotes the co-ordinate of the nucleon that absorbs the
neutrino. Then if the product $\vec {\delta P_k}  \cdot \vec X$ of the momentum
spread in the neutrino wave packet and the fluctuations in the position of the
detector  nucleon is small,  the exponential can be expanded and approximated
by the leading term

\beq{weak3}
\bra{f(E)}T \ket{i(E)} = 
\sum_k \bra{f(E)}T_o \cdot c_{k\mu} e^{i\vec {\delta P_k}  \cdot \vec X} \ket{i(E)} 
\approx
\sum_k \bra{f(E)}T_o \cdot c_{k\mu} \ket{i(E)}  
 \eeq

The transition matrix element for the probability that a muon is observed at
the detector is thus proportional to the coherent sum of the amplitudes
$c_{k\mu}$ for neutrino components with the same energy and different masses
and  momenta to produce a muon at the  detector. A similar result is obtained
for the probability of observing each other flavor. The final result is
obtained by summing the contributions over all the energies in the incident
neutrino wave packet. But as long as the flavor output for each energy is
essentially unchanged over the energy region in the wave packet, the flavor
ouput is already determined for each energy, and is independent of any
coherence or incoherence between components with different energies. 

For  the case of two neutrinos with energy $E$ and mass eigenstates $m_1$ and
$m_2$ the relative phase of the two neutrino waves at a distance $x$ is given
by eq. (\ref{shiftb}) with $E_1=E_2$

\beq{WW3a}
\phi^E_m(x)= (p_1 - p_2)\cdot x =
{{(p_1^2 - p_2^2)}\over{(p_1 + p_2)}}\cdot x  =
{{m_2^2-m_1^2}\over{2 \bar p}} \cdot x
\eeq

The flavor output of the detector is thus seen to be determined by the
interference between components in the neutrino wave paclet with the same
energy and different masses and momenta.   All the relevant physics is
in the initial state of the nucleon in the detector that detects the
neutrino and emits a charged lepton, together with the relative phases of
the components of the incident neutrino wave packet with the same energy.

This result (\ref{weak3}-\ref{WW3a}) is completely independent of the neutrino
source and in particular completely independent of whether the source satisfies
Stodolsky's stationarity condition\cite{Leo}. No subsequent time measurements
or additional final state interactions that mix energies can change this flavor
output result.

\section{Conclusions}

\subsection{Problems in calculating phase differences}

\begin{enumerate}

\item A reliable calculation of neutrino oscillations that avoids confusion 
requires an understanding of the effects of variations in the transit time of a
neutrino between source and detector and its quantum fluctuations. 

\item When the small  quantum fluctuations in time are neglected, the phase
difference between two components of a narrow wave packet with  different
masses depends only on their squared mass difference and the mean momentum in
the wave packet and is otherwise independent of their momenta and energies.

\item The value of this phase difference is given by the standard formula used
in all analyses of neutrino oscillations. 

\end{enumerate} 

\subsection{Problems in observing phase differences} 

\begin{enumerate} 

\item The observability of
this phase difference depends upon the quantum mechanics of the detector. 

\item
All coherence is destroyed in an ``ideal" detector, which precisely measures
the momentum and energy of the neutrino.  

\item A realistic detector preserves
the relative phase between states having the same energy and different momenta.
This is sufficient to uniquely determine the flavor output of the detector and
show that it satisfies the standard formula for neutrino oscillations. 

\item
Whether the coherence between states having different energies is destroyed, as
required by Stodolsky's stationarity theorem\cite{Leo}, or can somehow be
preserved by judicious time measurements is irrelevant to the flavor output of
the detector.   

\item The standard formula used in all analyses of neutrino
oscillations can be rigorously justified. All sources of confusion are resolved
by using states with the same energy and different momenta to calculate
neutrino oscillations. 

\end{enumerate}

\subsection{Some concluding remarks}
  
The initial uncertainty in the momentum of the detector nucleon in a localized
detector destroys all memory of the initial neutrino momentum and of the
initial neutrino mass after the neutrino has been absorbed.  The hand-waving
justification of the  result (\ref{weak3}) uses the uncertainty principle to
say that if we know where the detector is we don't know its momentum and can't
use momentum conservation to determine the mass of the incident neutrino. The
above rigorous justification shows full interference between the contributions
from different neutrino momentum states with the same energy as long as the
product of the momentum difference and the quantum fluctuations in the initial
position of the detector nucleon is negligibly small in the initial detector
state.   

This treatment of the neutrino detector is sufficient to determine the flavor
output of any experiment in which the incident neutrino wave packet is the same
well defined linear combination of mass eigenstates throughout the whole wave
packet. 

Time measurements and all possible coherence between amplitudes
from  components of the incident neutrino wave packet with different energies
are not considered and seen to be unnecessary. 
There may be fancy time measurements which
can introduce such coherence. But the coherence between incident neutrino
states with the same energy and different momentum already determines the
flavor output of the detector for each incident neutrino energy and cannot be
destroyed by  time measurements.

The question remains of possible variation of detector flavor output 
as a function of energy. As long as the wave packet is sufficiently
narrow in momentum this flavor output does not change appreciably over the
relevant energy range in the wave packet and the standard neutrino oscillation
formulas are valid. When the flavor output varies widely as a function of
energy, oscillations are no longer observed. This can be seen in the case of
neutrinos traveling large distances with many oscillation wave lengths, as in
neutrinos arriving from the sun or a supernova. Here the neutrino wave packet
can separate into components with different mass eigenstates, traveling with
different velocities and reaching the detector at measurably different times.
All this time variation appears simply\cite{Leo} in the energy spectrum, which
is the  fourier transform of the time behavior.

\section{acknowledgments}

It is a pleasure to thank Terry Goldman, Boris Kayser, Lev Okun, Georg Raffelt,
Alexei Smirnoff, Leo Stodolsky and Lincoln Wolfenstein for helpful discussions
and comments as well as to thank the organizers and participants in the
Ringberg CUTAPP 2005 (Leofest) workshop\cite{cutapp} for many clarifying
interactions. This work is supported in part by the U.S. Department of Energy,
Division of High Energy Physics, Contract W-31-109-ENG-38.

%
\catcode`\@=11 
\def\references{
\ifpreprintsty \vskip 10ex
%
\hbox to\hsize{\hss \large \refname \hss }\else
\vskip 24pt \hrule width\hsize \relax \vskip 1.6cm \fi \list
{\@biblabel {\arabic {enumiv}}}
{\labelwidth \WidestRefLabelThusFar \labelsep 4pt \leftmargin \labelwidth
\advance \leftmargin \labelsep \ifdim \baselinestretch pt>1 pt
\parsep 4pt\relax \else \parsep 0pt\relax \fi \itemsep \parsep \usecounter
{enumiv}\let \p@enumiv \@empty \def \theenumiv {\arabic {enumiv}}}
\let \newblock \relax \sloppy
 \clubpenalty 4000\widowpenalty 4000 \sfcode `\.=1000\relax \ifpreprintsty
\else \small \fi}
\catcode`\@=12 


\end{document}